\documentclass[11pt,a4paper]{article}
\usepackage{latexsym,amssymb,amsmath,amsthm,amsfonts,enumerate,verbatim,xspace,
exscale, setspace}

\input xy
\xyoption{all}
\CompileMatrices
\UseComputerModernTips

\usepackage{graphicx}
\usepackage{amsmath,amsbsy,amsfonts,amssymb}
 \usepackage{subfigure}

\usepackage[margin=1cm, 
font=small,format=plain,labelfont=bf,up,textfont=up]{caption}

\makeatletter
\def\blfootnote{\xdef\@thefnmark{}\@footnotetext}
\makeatother

\def\polhk#1{\setbox0=\hbox{#1}{\ooalign{\hidewidth\lower1.5ex\hbox{`}\hidewidth\crcr\unhbox0}}}

\title{Nonholonomic
LL systems on central extensions and  \\ the hydrodynamic Chaplygin sleigh with circulation}
\author{ Luis C. Garc\'{\i}a-Naranjo$^{a}$, Joris Vankerschaver$^{b, c}$}

\theoremstyle{plain}
\newtheorem{theorem}{Theorem}[section]

\newtheorem{proposition}[theorem]{Proposition}

\newtheorem*{theorem*}{Theorem}
\newtheorem{remarkth}[theorem]{Remark}
\theoremstyle{definition}

\newenvironment{remark}{\begin{remarkth}\upshape}{\hfill$\diamond$\end{remarkth}}

\def\vv<#1>{\langle#1\rangle}

\def\R{\mathbb{R}}
\def\I{\mathbb{I}}

\newcommand{\g}{\mathfrak{g}}
\def\se{\mathfrak{se}}

 \newcommand{\ad}{\mbox{$\text{\upshape{ad}}$}}

\newcommand{\vecu}{{\bf u}}
\newcommand{\In}{\mathcal{I}}
\newcommand{\Jn}{\mathcal{K}}

\def\M{\mathcal{M}}
\def\B{\mathcal{B}}

\begin{document}
\maketitle

\begin{abstract}
We consider nonholonomic systems whose configuration space is the central
extension of a Lie group and have  left invariant kinetic energy and constraints.
We study the structure of the associated Euler-Poincar\'e-Suslov equations
and show that there is a one-to-one correspondence between invariant measures on the original group and on the extended group. Our results are applied to the hydrodynamic Chaplygin sleigh,
that is, a planar rigid body that moves in a potential flow subject to a nonholonomic constraint modeling a fin or keel attached to the body, in the case where there is circulation
around the body.
\end{abstract}

\blfootnote{\noindent
\noindent

$^a$ Department de Matem\'atica Aplicada I, Universitat Politecnica de Catalunya, Barcelona, E-08028 Spain; email: luis.garcianaranjo@gmail.com\\

$^b$ Department of Mathematics, Imperial College London, SW7 2AZ, UK; e-mail: joris.vankerschaver@gmail.com\\
$^c$ Department of Mathematics, Ghent University, Krijgslaan 281, B-9000 Ghent, Belgium}

\tableofcontents

\section{Introduction and outline}

In this paper, we study the equations of motion for mechanical systems on central extension type Lie groups with nonholonomic constraints, where both the constraints and the kinetic energy are invariant under the left action of the group on itself. Our main motivating example comes from hydrodynamics and consists of a nonholonomic sleigh immersed in a two-dimensional potential flow with circulation. 

\paragraph{The Euler-Poincar\'e-Suslov equations.}

An {\em LL system} is a mechanical system on a Lie group $G$ with a kinetic energy Lagrangian and a set of nonholonomic constraints, so that both the Lagrangian and the constraints are left-invariant under the action of $G$ on itself.  Due to the invariance under the group action, the dynamics reduce to the Lie algebra $\g$, or to its dual $\g^*$ if working with the momentum formulation. The resulting reduced equations are termed the {\em Euler-Poincar\'e-Suslov} (EPS) equations \cite{FeKo1995}.

In this paper, we consider EPS equations associated to  nonholonomic LL 
systems for which the underlying Lie group is a \emph{central extension}.  We apply
 the criterion of Jovanovi\'c \cite{Jo1998} (see also \cite{Ko1988}) to obtain necessary and sufficient conditions on the existence
of invariant measures for these equations.   
One of our theoretical results is Theorem~\ref{T:inv-meas}, which states that an EPS system on a central extension $\widehat{G}$ has an invariant measure if and only if the corresponding system on the original Lie group $G$ has an invariant measure.

\paragraph{The hydrodynamic Chaplygin sleigh.}

 Our motivating example of an EPS system on a central extension is given by the motion of a two-dimensional rigid body which moves inside a potential flow with circulation $\kappa \ne 0$, where the nonholonomic constraint precludes motion transversal to the body, modeling, for instance, a very effective keel or fin. 

This model was first considered in the absence of circulation  in \cite{FeGa2010}, where
it was termed  the  {\em hydrodynamic Chaplygin sleigh}. This terminology reflects the fact that  in the absence of the fluid, the nonholonomic constraint models the effect of a sharp blade in the classical Chaplygin sleigh problem \cite{Ch2008} which prevents the sleigh from moving in the lateral direction. In the presence of the
fluid, the constraint can be interpreted as modeling the effect of a very effective keel or fin on the body \cite{FeGa2010}. It is an interesting historic coincidence that the name of Chaplygin is linked both to the development of the Chaplygin-Lamb equations \cite{Ch1933} as well as to the nonholonomic Chaplygin sleigh \cite{Ch2008}.  Similar models for two-dimensional swimmers have been studied in \cite{KeHu2006} (see also \cite{RaRa2000}).  The motion of the hydrodynamic Chaplygin sleigh in the presence of circulation is treated in \cite{FeGaVa2012}. However, to the best of our knowledge this is the first time that the geometric nature of the system is elucidated.

\paragraph{The Chaplygin-Lamb equations.}

When the effect of the keel is ignored, so that there are no nonholonomic constraints, the equations of motion for the hydrodynamic sleigh reduce to the Chaplygin-Lamb equations \cite{Ch1933, La1945}. We show that these equations can be viewed in two different, but equivalent ways: 
\begin{enumerate}
	\item As a left-invariant system on the group $\operatorname{SE}(2)$ of translations and rotations in the plane, moving under the influence of a \emph{gyroscopic force}. The latter is termed the \emph{Kutta-Zhukowski force} \cite{Mi1968} and models the effect of nonvanishing circulation on the body.
	
	\item As a geodesic system on a \emph{central extension} of $\operatorname{SE}(2)$ by $\R^3$ that we denote by $\widehat{G}$, where the extra variables in the $\R^3$-factor describe the circulation. In this way, the Kutta-Zhukowski force becomes a geometric effect, which is not added explicitly to the system but appears a posteriori as a consequence of how the central extension is constructed. 
\end{enumerate}

A classical counterpart of this duality is the description of a particle of charge $e$ moving under the influence of a magnetic field $B$ perpendicular to the plane of motion. As is well known, such a particle may be modeled either as moving under the influence of the Lorentz force, or as a particle moving in the Heisenberg group $\mathbb{R}^2 \times \mathbb{S}^1$ equipped with a group multiplication involving the magnetic field (see \cite{Mo2002}).   In the example of the hydrodynamic Chaplygin sleigh, the circulation $\kappa$ plays the role of the charge, and the cocycle will be discussed below.

\paragraph{The nature of the cocycles.}

In constructing the extension $\widehat{G}$ of $\operatorname{SE}(2)$ by $\R^3$, we introduce an $\R^3$-valued two-cocycle $C: \mathfrak{se}(2) \times \mathfrak{se}(2) \to \R^3$ which can be decomposed on fluid-dynamical grounds as $C = (C_1, C_2)$, where $C_1$ takes values in $\R$ while $C_2$ is $\R^2$-valued.  As we have pointed out before, each of these cocycles is responsible for the appearance of certain gyroscopic forces in the equations of motion, and we now discuss these forces some more.

The first cocycle, $C_1$, is ``essential'' in the sense that it cannot be written as the coboundary (defined below) of a one-cocycle $\mathcal{A}$, and we argue that this is a consequence of Kelvin's theorem, which states that circulation is constant.  By contrast, the second cocycle $C_2$ is exact, and we show that it can be ``gauged away'' by adequately choosing the origin of  the body reference frame. From a physical point of view, the cocycle $C_2$ is associated to the moment generated by the Kutta-Zhukowski force.  While it would have been possible to get rid of $C_2$, this would complicate the description of the nonholonomic constraint that we discuss below.

\paragraph{Adding nonholonomic constraints.}

The effect of the keel gives rise to a nonholonomic constraint on the system, which can be viewed as follows: if we affix a frame $\{ \mathbf{E}_1, \mathbf{E}_2 \}$ to the body, with $\mathbf{E}_1$ aligned with the keel and $\mathbf{E}_2$ perpendicular to it, the effect of the keel is to preclude motion in the $\mathbf{E}_2$-direction, or in other words
\begin{equation} \label{eq:intro_nh}
	v_2 = 0,
\end{equation}
where $v_2$ is the component of the body velocity in the direction of $\mathbf{E}_2$.
This is a constraint on the velocities which cannot be integrated to give a relation between the admissible configurations of the body, and is therefore nonholonomic.  Just as the kinetic energy, this constraint is left invariant under the action of the central extension $\widehat{G}$ on itself, and therefore gives rise to an EPS system on $\widehat{\mathfrak{g}}$, the Lie algebra of $\widehat{G}$.

Using the geometric structure of the equations, we are able to obtain necessary and sufficient conditions for the existence of an invariant measure (Proposition~\ref{prop:chaplygin_measure}). Among other things, we show that the existence of an invariant measure is independent of the circulation.

\paragraph{Outline.} The paper is organized as follows. In Section \ref{Sec:Central-Extension}
we recall the necessary preliminaries on the theory of central extensions of Lie groups with the viewpoint
on their mechanical applications. In Section \ref{S:LL-central-ext} we consider the EPS equations
of LL systems whose underlying Lie group is a central extension and give conditions for the existence
of invariant measures in Theorem \ref{T:inv-meas}. 

 Section \ref{sec:nonhchaplygin} considers the structure of the equations of motion for planar rigid  bodies
 moving on a perfect fluid with circulation. Our contributions are contained in Theorem \ref{T:Lie-Poisson_structure_of_Chap_Lamb} that describes the geometric structure of the most general form of the Chaplygin-Lamb equations
 and in the construction in subsection \ref{SS:hydro-Chap-sleigh} where we show that
 the reduced equations for the hydrodynamic Chaplygin sleigh with circulation are of EPS type,
 and where we give conditions for measure preservation.
 These results rely on the introduction of the  group cocycle that defines a central extension of the
  group $\operatorname{SE}(2)$ described above, and  that  is studied in detail in Section \ref{A:SE(2)-Extensions}.

\section{Central extensions of Lie groups}
\label{Sec:Central-Extension}

We briefly recall the definitions and basic properties of central extensions of Lie groups to introduce
the relevant notation. A detailed account of the geometry of central extensions in the context of mechanics can be found in  \cite{MaMiOrPeRa2007, LiMa1987, KhWe2009}.

\paragraph{Definition.}
Let $G$ be a Lie group and $A$ an abelian Lie group. We will use additive notation for the group operation in $A$.  For our purposes, a \emph{central extension} of $G$ by $A$ is a Lie group $\widehat{G}$ such that $\widehat{G} = G \times A$ equipped with the group multiplication
\begin{equation} \label{central_multiplication}
(g,\alpha)(h,\beta)=(gh, \alpha + \beta + B(g,h)),
\end{equation}
where $B:G\times G\to A$ is
a normalized group \emph{two-cocycle}. 
Associativity of the multiplication on $\widehat G$ is equivalent to the two-cocycle identity,
\begin{equation*}
\label{E:Cocycle_identity}
B(f,g)+B(fg,h)=B(f,gh)+B(g,h) \qquad \mbox{for all} \qquad f,g,h\in G.
\end{equation*}
The assumption that the two-cocycle $B$ is normalized can be made without loss of generality and
amounts to
\begin{equation*}
B(g,e)=B(e,g)=0 \qquad \mbox{for all} \qquad g\in G,
\end{equation*}
 implying that  $B(g,g^{-1})= B(g^{-1},g)$.

\paragraph{Central extension of Lie algebras.}
The Lie algebra $\widehat \g$ of $\widehat{G}$ is isomorphic as a vector space to $\g\times\mathfrak{a}$, and is equipped with the following bracket: 
\begin{equation*} \label{CAbracket}
[(\xi, a), (\eta,  b)]_{\widehat \g} =([\xi,\eta]_{\g}, C(\xi,\eta))
\end{equation*}
for $(\xi, a), \,(\eta,  b)\in \widehat \g$.   Here, the $\mathfrak{a}$-valued Lie algebra two-cocycle $C:\g\times \g \to \mathfrak{a}$ is defined by
\begin{equation}
\label{E:Lie_algebra_cocycle}
C(\xi, \eta) := \left .\frac{\partial ^2}{\partial s \partial t} \right |_{s=t=0} (B(g(t) ,h(s))-B(h(s),g(t))), \qquad i=1,2,
\end{equation}
with $g(t), \, h(s)$ smooth curves on $G$ satisfying $\dot g(0)=\xi, \; \dot h(0)=\eta$.

It may happen that $C$ can be written in terms of a one-cocycle $\mathcal{A}: \g \to \mathfrak{a}$ as $C(\xi, \eta) = - \mathcal{A}([\xi, \eta])$ for all $\xi, \eta \in \g$.  In this case, $C$ is a \emph{coboundary} in the sense of Lie algebra cohomology, and we write $C = \delta \mathcal{A}$.   When this happens, the central extension is said to be \emph{trivial}, as the mapping 
\begin{equation} \label{trivmapping}
	\Psi_\mathcal{A}: \g \times \mathfrak{a} \to \widehat \g, \quad
	(\xi, a) \mapsto (\xi, a - \mathcal{A}(\xi))
\end{equation}
 then determines a Lie algebra isomorphism between the Lie algebra $\g \times \mathfrak{a}$ with the product bracket and the central extension $\widehat \g$.

\paragraph{Lie-Poisson structures.}
As a vector space, the dual Lie algebra $\widehat \g^*$ equals $\g^*\times\mathfrak{a}^*$. For $(\mu, \sigma) \in \widehat \g^*$, the $\pm$ Lie-Poisson 
bracket of functions $F,K\in C^\infty( \widehat \g^*)$ is readily computed to be
\begin{equation}
\label{E:Lie_Poisson_C_Ext}
\{F,K\}_{\widehat \g^*}^\pm(\mu, \sigma) =\pm \left \langle \mu , \left  [\frac{\delta F}{\delta \mu} , \frac{\delta K}{\delta \mu}  \right ] \right \rangle\pm  \left \langle \sigma
 , C\left  (\frac{\delta F}{\delta \mu} , \frac{\delta K}{\delta \mu}  \right ) \right \rangle.
\end{equation}

Notice that this bracket only involves functional derivatives with respect to $\mu$ so the 
components of $\sigma$ are Casimir functions.  Therefore, if we fix the 
value of $\sigma \in \mathfrak{a}^\ast$ in \eqref{E:Lie_Poisson_C_Ext}, we obtain a non-canonical 
Poisson bracket on $\mathfrak{g}^\ast$, given by formally the same expression
 as \eqref{E:Lie_Poisson_C_Ext}: for $f, k \in C^\infty(\mathfrak{g}^\ast)$ we have 
\begin{equation} \label{noncanpoisson}
\{f, k\}^\pm_\sigma(\mu) =\pm \left \langle \mu , \left  [\frac{\delta f}{\delta \mu} , \frac{\delta k}{\delta \mu}  \right ] \right \rangle\pm  \left \langle \sigma
 , C\left  (\frac{\delta f}{\delta \mu} , \frac{\delta k}{\delta \mu}  \right ) \right \rangle,
\end{equation}
where $\sigma \in \mathfrak{a}^\ast$ is now regarded as fixed.  Roughly speaking, we therefore have a one-to-one correspondence between Poisson brackets on $\mathfrak{g}^\ast$ which are the sum of a Lie-Poisson term and a cocycle, and Lie-Poisson brackets on central extensions $\widehat \g^*$.  This can be made rigorous by observing that the injection $\iota_\sigma : \mathfrak{g}^\ast \hookrightarrow \widehat \g^*$, given by $\iota_\sigma(\mu) := (\mu, \sigma)$ for $\sigma$ fixed, is a Poisson map taking the non-canonical Poisson structure \eqref{noncanpoisson} into the Lie-Poisson structure \eqref{E:Lie_Poisson_C_Ext}.

In the case of a trivial central extension, the cocycle term in \eqref{noncanpoisson} can be ``gauged away'': let $C = \delta \mathcal{A}$.  For $\sigma \in \mathfrak{a}^\ast$, we denote by $\mathcal{A}_\sigma: \g \to \mathbb{R}$ the linear map defined by $\mathcal{A}_\sigma(\xi) := \left< \sigma, \mathcal{A} (\xi) \right>$ for all $\xi \in \g$.  Note that $\mathcal{A}_\sigma \in \g^\ast$.  The Poisson bracket \eqref{E:Lie_Poisson_C_Ext} can then be rewritten as 
\begin{align*}
	\{F,K\}_{\widehat \g^*}^\pm(\mu, \sigma) & = \pm \left \langle \mu , \left  [\frac{\delta F}{\delta \mu} , \frac{\delta K}{\delta \mu}  \right ] \right \rangle \mp \left \langle \sigma
 , \mathcal{A} \left( \left[\frac{\delta F}{\delta \mu} , \frac{\delta K}{\delta \mu} \right] \right ) \right \rangle \\
 & = \pm \left \langle \mu - \mathcal{A}_\sigma, \left  [\frac{\delta F}{\delta \mu} , \frac{\delta K}{\delta \mu}  \right ] \right \rangle, 
\end{align*}
so that the cocycle vanishes apart from a shift $\mu \mapsto \mu - \mathcal{A}_\sigma$ in the momenta of the system.   

\begin{proposition}
The shift map $\Phi_\mathcal{A} : \widehat \g^\ast \to \g^\ast \times \mathfrak{a}^\ast$, given by $\Phi_{\mathcal{A}}(\mu, \sigma) = (\mu - \mathcal{A}_\sigma, \sigma)$ is the dual of the trivialization mapping \eqref{trivmapping}.  Moreover, the shift map  is a Poisson map, taking the Lie-Poisson structure \eqref{E:Lie_Poisson_C_Ext} on $\widehat \g^\ast$ into the product Lie-Poisson structure on $\g^\ast \times \mathfrak{a}^\ast$. 
\end{proposition}

\paragraph{Hamiltonian vector fields.}
The Hamiltonian vector field of a function $H$ on $ \widehat \g^*$ is defined by the equation
\begin{equation}
\label{E:HamVF-central}
\left ( \dot \mu , \dot \sigma \right ) = \mp
 \widehat{ \ad^*}_{ \left ( \frac{\delta H}{\delta \mu} , \frac{\delta H}{\delta \sigma}  \right ) } (\mu, \sigma)
= \left ( \mp \left (  \ad^*_{\frac{\delta H}{\delta \mu}} \mu + \sigma \circ C \left (\frac{\delta H}{\delta \mu}, \cdot  \right ) \right ) \, , \,   0 \right ).
\end{equation}
where $\widehat{ \ad^*}$  (respectively, ${ \ad^*}$) denotes the infinitesimal coadjoint action
on $\widehat{ \g}^*$ (respectively,  
on   $\g^*$).
Notice that $\sigma$ is constant throughout the motion as expected.  

A well-known but instructive example of a mechanical system on a central extension is given by the motion of a charged particle under the influence of a constant magnetic field $B$ perpendicular to the plane of motion (see \cite{Mo2002}). We assume that the motion takes place in the $xy$-plane, while $\mathbf{B} = B \mathbf{e}_z$ is parallel to the $z$-axis. For this example the group $G$ is $\mathbb{R}^2$, the Abelian group $A$ is $U(1) \cong \mathbb{S}^1$, and the magnetic field defines a cocycle on $\mathbb{R}^2$ with values in $\mathbb{R} \cong \mathfrak{u}(1)$ given by 
\[
	C(\mathbf{x}, \mathbf{y}) = \mathbf{B} \cdot( \mathbf{x} \times \mathbf{y}),
\]
for all $\mathbf{x}, \mathbf{y} \in \mathbb{R}^2$. The central extension obtained in this way is the \emph{Heisenberg group}. The Lie algebra of this group is $\mathfrak{h} \cong \mathbb{R}^2 \times \mathbb{R}$, equipped with the Lie bracket $[(\mathbf{v}, \alpha), (\mathbf{w}, \beta) ] = (\mathbf{0}, C(\mathbf{v}, \mathbf{w}))$.  For the equations of motion \eqref{E:HamVF-central} we then have that the $\mathrm{ad}^\ast$-term on the right hand side vanishes, since $\mathbb{R}^2$ is Abelian. If we denote $\mu$ as $\mathbf{p} = (p_x, p_y)$ and let $\mathbf{v} := \delta H/\delta \mu$, we have that \eqref{E:HamVF-central}, with the minus sign, becomes 
\[
	\dot{\mathbf{p}} = \sigma \mathbf{v} \times \mathbf{B}, 
		\quad \text{and} \quad
	\dot{\sigma} = 0, 
\]	
where the former can be integrated to $\sigma = e$, a constant. This shows that the canonical equations of motion on the Heisenberg group give rise to the familiar Lorentz equations of electrodynamics. This example shows that the effect of having the cocycle in the equations \eqref{E:HamVF-central} is to add a \emph{gyroscopic force} to the system (i.e. a force which is at right angles to the velocity).  We will extend this observation to the dynamics of a rigid body moving under the influence of the Kutta-Zhukowski force in Section~\ref{A:SE(2)-Extensions}.

\section{Nonholonomic LL systems on central extensions}
\label{S:LL-central-ext}

To the best of our knowledge, the derivation of the \emph{Euler--Poincar\'e--Suslov (EPS) equations} has never been made explicit in the case where the underlying Lie group is a central extension. Here we develop the general theory in detail.  In section~\ref{sec:nonhchaplygin}, we apply this theory to the hydrodynamic Chaplygin sleigh with circulation.

\subsection{Euler-Poincar\'e-Suslov equations on a Lie group $G$.} 
In general, a nonholonomic system on a Lie group $G$ with a left invariant kinetic energy Lagrangian and
left invariant constraints is termed an \emph{LL system}.
Due to invariance, the dynamics reduce to the Lie algebra $\g$, or to its dual $\g^*$ if working with the momentum formulation.  We start from a reduced Lagrangian $L:\g\rightarrow \R$, which defines an inertia operator $\I:\g \rightarrow \g^*$ by the
relation
\begin{equation*}
L(\xi)= \frac{1}{2}\langle \I\xi , \xi \rangle, \qquad \mbox{for} \;\; \xi \in \g,
\end{equation*}
where $\langle \cdot , \cdot \rangle$ denotes the duality pairing. The reduced Hamiltonian, $H:\g^*\rightarrow \R$,
is then given by
\begin{equation*}
H(\mu)= \frac{1}{2}\langle \mu , \I^{-1} \mu \rangle, \qquad \mbox{for} \;\; \mu \in \g^*.
\end{equation*}

The nonholonomic constraints are expressed in terms of $n$ linearly independent fixed covectors $\nu_i\in \g^*$, $i = 1, \ldots, n$: we say that an instantaneous velocity $\xi\in \g$ satisfies the constraints if
\begin{equation}
\label{E:LL-constraints}
\langle \nu_i , \xi \rangle =0, \qquad i=1, \dots, n.
\end{equation}
We let $\mathcal{D} \subset \mathfrak{g}$ be the vector subspace of all velocities satisfying the constraints, and we say that the constraints are nonholonomic if $\mathcal{D}$ is not a Lie subalgebra of $\g$.

The reduced EPS equations on $\g^*$  are given by, see e.g. \cite{Bl2003},
\begin{equation}
\label{E:Euler-Poincare-Suslov}
\dot \mu=\ad^*_{\I^{-1}\mu} \mu + \sum_{i=1}^n\lambda_i \nu_i,
\end{equation}
where the multipliers $\lambda_i, \, i=1,\dots, n$, are certain scalars that are uniquely determined by the condition that the constraints (\ref{E:LL-constraints})
are satisfied.  Explicitly, the Lagrange multipliers are given by 
\begin{equation}
\label{E:Lagrange_multiplier}
\lambda_j=-\sum_{i=1}^n (D^{-1})_{ij}\langle \ad^*_{\I^{-1}\mu} \mu, \I^{-1}\nu_i \rangle,
\end{equation}
where $D$ is the matrix with components $D_{ij} := \langle \nu_i,\I^{-1}\nu_j \rangle, \; i,j=1,\dots n$, and $D^{-1}$ is its inverse.

\subsection{Euler-Poincar\'e-Suslov equations on central extensions.}

Now suppose that  $\widehat G$ is  a central extension of $G$ by the abelian Lie group $A$
as explained in Section~\ref{Sec:Central-Extension}.  We assume that $\widehat G \cong G \times A$ as a manifold and with the multiplication given by \eqref{central_multiplication}.  We let $L$ be a left-invariant Lagrangian on $G$, with associated Hamiltonian $H$, and we let $\nu_i\in \g^\ast$, $i = 1, \ldots, n$ be a set of linearly independent constraint covectors.  We now wish to ``lift'' these data to the central extension $\widehat{G}$, so that we can derive the corresponding EPS equations on the co-algebra $\widehat \g^*$.

By left translating the co-vectors  $\nu_i\in \g^*$ one can define
left-invariant constraint one-forms $\epsilon_i,$ on $G$, given by $\epsilon_i(g)=T_g^*L_{g^{-1}}\nu_i \in T_g^*G, \;  i=1, \dots, n$.   Since $\widehat G \cong G \times A$, these constraint one-forms naturally induce constraint one-forms $\widehat \epsilon_i$ on $\widehat G$, given by $\widehat \epsilon_i(g, \alpha) = (\epsilon(g), 0)$.  Likewise, the co-vectors $\nu_i \in \g^*$ can be lifted to co-vectors $\widehat \nu_i = (\nu_i, 0)$ in $\widehat{\g}^\ast$, and we have that $\widehat \epsilon_i(g, \alpha) = T^\ast_{(g, \alpha)} L_{(g, \alpha)^{-1}} \widehat{\nu}_i$.  

Secondly, we define the left invariant, kinetic energy Hamiltonian
 $H_{T^*\widehat G}:T^*\widehat G \to \R$, whose value
at the identity is given by
\begin{equation}
\label{E:label_Ham_ext}
H_{\widehat \g^*}(\mu, \sigma) = H(\mu)+ \frac{1}{2}||\sigma||^2, \qquad
\mbox{for} \; (\mu, \sigma)\in \g^*\times \mathfrak{a}^*=\widehat{\g}^*,
\end{equation}
where $||\cdot ||^2$ denotes  any positive definite, quadratic form on $\mathfrak{a}^*$. For convenience
we write
\begin{equation*}
H_{\widehat \g^*}(\mu, \sigma) = \left \langle \, (\mu,\sigma ) \,  ,  \, \widehat \I^{-1} (\mu,\sigma ) \,
\right \rangle
\end{equation*}
where the non-degenerate, extended, inertia tensor 
$\widehat \I:\widehat{\g}\to \widehat{\g}^*$ is determined from \eqref{E:label_Ham_ext}. As we shall see, the choice of the quadratic form  $||\cdot ||^2$  on $\mathfrak{a}^*$ does not affect the final form of the equations.

We have thus extended the left invariant constraints and kinetic energy on $G$ to 
define an LL system on the central extension $\widehat G$. The corresponding
EPS equations  \eqref{E:Euler-Poincare-Suslov} become,
\begin{equation*}
\begin{split}
(\dot \mu,\dot\sigma) &=  \widehat{ \ad^*}_{\widehat{\I}^{-1}(\mu, \sigma)} (\mu, \sigma) + \sum_{i=1}^n\lambda_i \widehat \nu_i \\
&= ( \ad^*_{\I^{-1}\mu} \mu +\sigma \circ C( \I^{-1}\mu, \cdot ) , 0 ) +
\sum_{i=1}^n\lambda_i ( \nu_i , 0),
\end{split}
\end{equation*}
where we have used \eqref{E:HamVF-central}. Here, the  
Lie algebra two-cocycle $C:\g\times \g \to \mathfrak{a}$ is defined by
\eqref{E:Lie_algebra_cocycle}.

The above equations can be written as
\begin{equation}
\label{E:Euler-Poincare-Suslov-Cent-Ext}
\begin{split}
\dot \mu=\ad^*_{\I^{-1}\mu} \mu +\sigma \circ C( \I^{-1}\mu ,\, \cdot \,  ) +\sum_{i=1}^n\lambda_i \nu_i, \qquad 
\dot \sigma =0,
\end{split}
\end{equation}
and one finds the following expression for the multipliers:
\begin{equation}
\label{E:Lagrange_multiplier_ext}
\lambda_j=-\sum_{i=1}^n (D^{-1})_{ij}\left ( \langle \ad^*_{\I^{-1}\mu} \mu, \I^{-1}\nu_i \rangle + 
\langle  \sigma , C( \I^{-1}\mu ,\, \I^{-1}\nu_i) \rangle \, \right  ).
\end{equation}
Notice that these equations are independent of the choice of the quadratic form  $||\cdot ||^2$  on $\mathfrak{a}^*$ as advertised, and reduce to \eqref{E:Euler-Poincare-Suslov}, \eqref{E:Lagrange_multiplier} 
if $\sigma=0$.

\subsection{Existence of invariant measures.} 

It is natural to ask whether the equations 
\eqref{E:Euler-Poincare-Suslov-Cent-Ext} and \eqref{E:Lagrange_multiplier_ext} possess an invariant
measure. We answer this question using the criterion of Jovanovi\'c \cite{Jo1998} for the existence of an invariant measure for the EPS equations on the Lie algebra of an arbitrary Lie group.  We assume that there is only one constraint, so that  $n=1$; the case of multiple constraints can be dealt with in a similar way.
Following \cite{Jo1998}, the necessary and sufficient condition for equations \eqref{E:LL-constraints}, \eqref{E:Euler-Poincare-Suslov}
to have an invariant measure  is that the constraint covector 
$\widehat{\nu}=\widehat{\nu}_1 \in \widehat{\g}^*$ satisfies
\begin{equation}
\label{E:inv-meas}
\frac{1}{\langle \widehat{\nu} ,\widehat{ \I}^{-1}\widehat{\nu} \rangle} \, 
\widehat{\operatorname{ad}^*}_{\widehat{\I}^{-1}\widehat{\nu}}\widehat{\nu} + \widehat{T}
 = c\widehat{\nu}, \qquad c\in \R,
\end{equation}
where $\widehat{T}\in \widehat{\g}^*$ is defined by the relation $\langle \widehat{T}, \widehat{\xi}
 \rangle = \operatorname{trace}(\widehat{\operatorname{ad}}_{\widehat{\xi}})$, 
 $\widehat{\xi}\in \widehat{\g}$.

However, for $(\xi, a), (\eta, b)\in \g\times \mathfrak{a}=\widehat{\g}$ we have 
\begin{equation*}
\widehat{\operatorname{ad}}_{(\xi,a)}(\eta, b) =(\operatorname{ad}_\xi\eta , C(\xi, \eta)).
\end{equation*}
It follows that the  operator 
$\widehat{\operatorname{ad}}_{(\xi,a)}: \widehat{\g}=\g \times \mathfrak{a} \to 
\widehat{\g}=\g \times \mathfrak{a}$ has matrix representation 
\begin{equation*}
\left ( \begin{array}{cc} \operatorname{ad}_\xi & 0 \\ C(\xi, \cdot) & 0 \end{array} \right ).
\end{equation*}
Hence $\operatorname{trace}(\widehat{\operatorname{ad}}_{(\xi,a)})=\operatorname{trace}({\operatorname{ad}}_{\xi})$ and we can write
$
\widehat T = (T,0)\in \g^*\times \mathfrak{a}^*, 
$
where $T\in {\g}^*$ is defined by  $\langle {T}, {\xi}
 \rangle = \operatorname{trace}({\operatorname{ad}}_{{\xi}})$ for $\xi \in \g$. In addition,
 since $\widehat{\nu}=(\nu,0)$, we can write \eqref{E:inv-meas} in components as
 \begin{equation*}
\frac{1}{\langle \nu , \I^{-1}{\nu} \rangle} \, 
{\operatorname{ad}^*}_{({\I}^{-1}{\nu},0)}({\nu},0) + ({T},0)
 = c({\nu},0), \qquad c\in \R.
\end{equation*}
Therefore, the condition \eqref{E:inv-meas} is equivalent to 
 \begin{equation*}
\frac{1}{\langle \nu , \I^{-1}{\nu} \rangle} \, 
{\operatorname{ad}^*}_{{\I}^{-1}{\nu}}{\nu} + {T}
 = c{\nu}, \qquad c\in \R,
\end{equation*}
which is precisely the necessary and sufficient condition for the equations
\eqref{E:Euler-Poincare-Suslov}, \eqref{E:Lagrange_multiplier} to possess an
invariant measure. This analysis can be generalized to the case where the number
$n$ of constraints is arbitrary. This shows:
\begin{theorem}
\label{T:inv-meas}
The Euler-Poincar\'e-Suslov equations \eqref{E:Euler-Poincare-Suslov-Cent-Ext} and \eqref{E:Lagrange_multiplier_ext} on the dual Lie algebra $\widehat{\g}^*$ of the central extension
$\widehat G$ of the Lie group $G$, possess an invariant measure for an arbitrary value of
$\sigma \in \mathfrak{a}^*$ if and only if they possess an invariant measure for the specific
value $\sigma =0$. In other words, such an invariant measure exists if and only if the 
Euler-Poincar\'e-Suslov equations
 \eqref{E:Euler-Poincare-Suslov}, \eqref{E:Lagrange_multiplier} on ${\g}^*$ possess an
 invariant measure.
\end{theorem}

\section{The motion of a planar rigid body in a perfect fluid}
\label{sec:nonhchaplygin}

We will now present a mechanical example that fits the geometric construction given in section \ref{S:LL-central-ext}.
This example concerns 
the generalization of the  hydrodynamic version of the Chaplygin sleigh treated in \cite{FeGa2010} to the case when there is circulation around the body.

Most of the material in sections \ref{sec:kinematics} and \ref{sec:dynamics} ahead contain the preliminaries necessary to treat our problem and  can be found, for instance, in \cite{No1981, KaMaRoMe2005} as well as in the classical works of Lamb \cite{La1945} and Milne-Thomson \cite{Mi1968}.
  However we reach out to give an original result in Theorem \ref{T:Lie-Poisson_structure_of_Chap_Lamb} that describes the geometric structure of the most general
version of the Chaplygin-Lamb equations.  The treatment of the hydrodynamic Chaplygin sleigh with circulation is 
presented in section \ref{SS:hydro-Chap-sleigh}.

\subsection{Kinematics }
\label{sec:kinematics}

\begin{figure}[h]
\centering
\subfigure[Body frame is aligned with axes of symmetry of the body.]{\includegraphics[width=5cm]{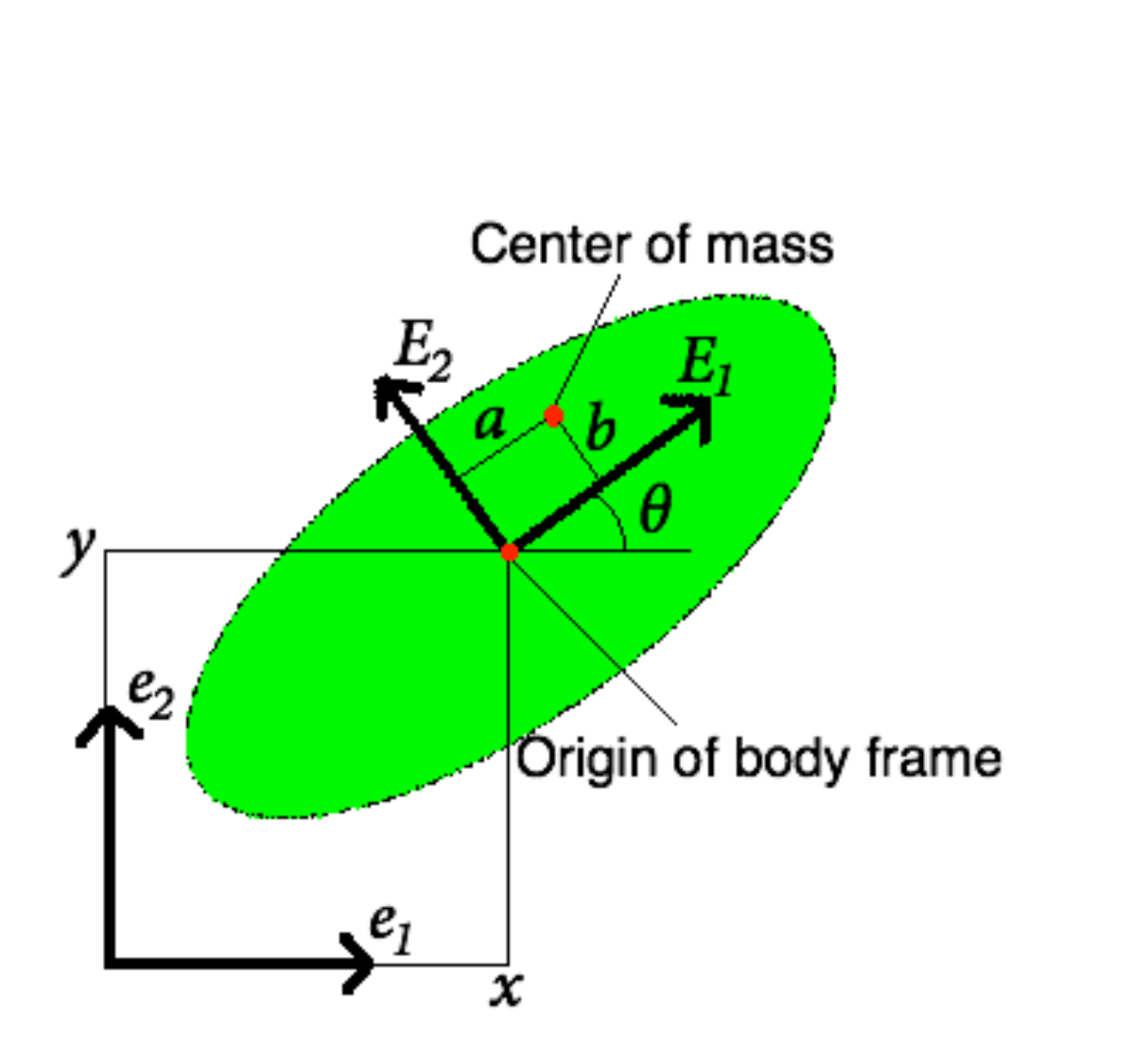}} \qquad \qquad
\subfigure[Arbitrary position and orientation of the body frame.]{\includegraphics[width=5cm]{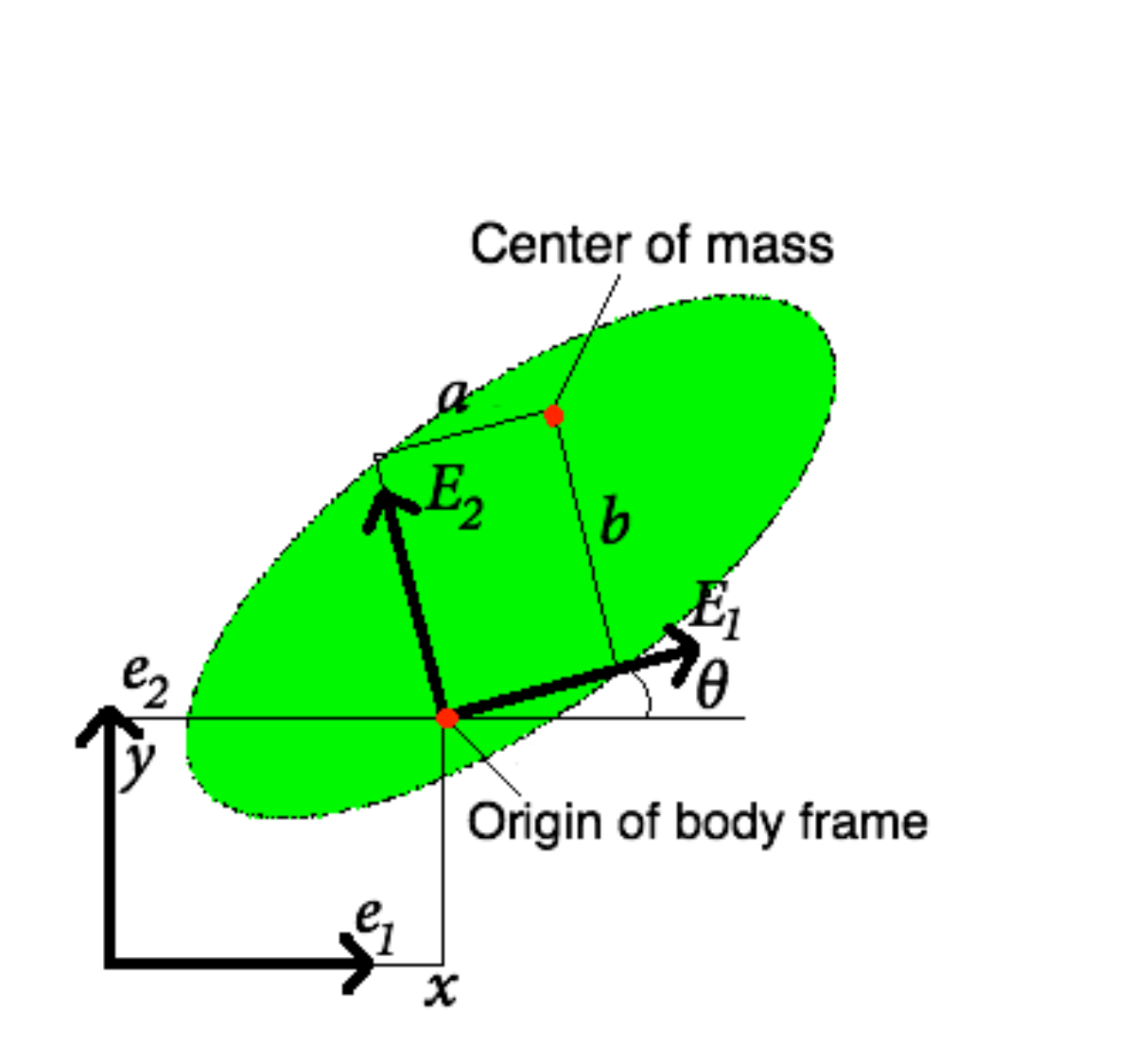}} 
\caption{
Two different choices of the body  frame for an elliptical  two-dimensional rigid body. In both cases the origin of
the body frame does not coincide with the center of mass. \label{F:kinematics-diagram}
}
\end{figure}

We adopt Euler's approach to the study of the rigid body dynamics and consider an orthonormal \emph{body frame}
$\{{\bf E}_1, {\bf E}_2\}$ that is attached to the body. This frame is related to a fixed \emph{space frame}
$\{{\bf e}_1, {\bf e}_2\}$
by a rotation  by an angle $\theta$ that specifies the orientation of the two dimensional body at each time.
We will denote by ${\bf x}=(x, y) \in \R^2$ the spatial coordinates of the origin of the body frame  and we do not assume that the origin of the body frame is located at the center
of mass. We will denote by $(a,b)$  the (constant) coordinates of the center of mass in the body frame
(see Figure 
\ref{F:kinematics-diagram}).
The configuration of the body at any time is  
completely determined by the element  $g$ of the two dimensional Euclidean group  
$\operatorname{\operatorname{SE}}(2)$ given by
\begin{equation*}
g= \left ( \begin{array}{ccc} \cos \theta  & - \sin \theta  & x \\ \sin \theta  & \cos \theta  & y \\ 0 & 0 & 1 \end{array} \right ) \in \operatorname{\operatorname{SE}}(2). 
\end{equation*}

We will often denote the  above element in $g\in\operatorname{\operatorname{SE}}(2)$ by $g=(R_{\theta}, {\bf x})$, where
$R_{\theta} \in \operatorname{SO}(2)$ is the  rotation matrix determined by the angle $\theta$.
 Let $(v_1, v_2)\in \R^2$ be the linear velocity of the origin of the body frame written in 
the body coordinates, and denote by $\omega  = \dot \theta $ the body's angular velocity. They define the element $\xi $ in the Lie algebra $\se(2)$
given by 
\begin{equation} \label{relgroupalgebra}
  \xi = g^{-1}\dot g= \left ( \begin{array}{ccc} 0 & - \omega   & v_1 \\ \omega  & 0& v_2 \\ 0 & 0 & 0 \end{array} \right ) \in \se (2).
\end{equation}
Explicitly we have
\begin{equation}
\label{E:Reconstruction_Equations_2D}
\dot \theta = \omega, \qquad v_1=\dot x \cos \theta  +  \dot y\sin \theta , \quad v_2= -\dot x\sin \theta   +  \dot y \cos \theta .
\end{equation}
For convenience, we will sometimes identify $\se (2)$ with $\R^3$ as vector spaces, and denote $\xi \in \se(2)$ as the column vector $(\omega, v_1, v_2)^T\in \R^3$. The Lie algebra commutator
takes the form
\begin{equation*}
	  \left [ \, (\omega, v_1,v_2) \, , \,  (\omega', v_1',v_2')\, \right ]_{\mathfrak{se}(2)} 	  = ( \, 0 \, , \,  v_2\omega'-\omega v_2' \, , \,  \omega v_1' -v_1 \omega ' \,  ).
\end{equation*}

For future reference, we give an explicit description of the dual space $\se(2)^\ast$.  Since $\se(2)$ is isomorphic to $\mathbb{R}^3$ and using the Euclidian inner product, we have that $\se(2)^\ast \cong \mathbb{R}^3$.  A typical element $\mu$ is represented as a row vector $\mu = (k, p_1, p_2)$.  The duality pairing between $\mu$ and an element $\xi= (\omega, v_1, v_2)^T$ of $\se(2)$ is given by 
\begin{equation*}
\label{E:pairing_in_se(3)}
\langle \mu, \xi \rangle =\mu \xi=k\omega +p_1v_1+p_2v_2.
\end{equation*}

The dual space $\se(2)^\ast$ is equipped with the (minus) \emph{Lie-Poisson bracket}, which is given by 
\[
	\{ F, K \}^-_{\mathfrak{se}(2)^\ast}(\mu)  =- \left< \mu, \left[ \frac{\delta F}{\delta \mu}, \frac{\delta K}{\delta \mu} \right] \right>
\]
for all functions $F, K$ on $\mathfrak{se}(2)^\ast$.  In coordinates, we have 
\begin{equation}
\label{E:Lie-Poisson-se(2)}
	\{ F, K \}^-_{\mathfrak{se}(2)^\ast}(\mu) = ( \nabla_\mu F)^T 
		\begin{pmatrix}
		    0 & -p_2 & p_1 \\
		    p_2 & 0 & 0 \\
		    -p_1 & 0 & 0 
		\end{pmatrix}
		(\nabla_\mu K),
\end{equation}
where $\nabla_\mu F$ is the gradient of $F$ with respect to the variables $(k, p_1, p_2) = \mu$.

\paragraph{The fluid flow at a given instant.} Consider now the motion of the fluid that
surrounds the body. Suppose that at a given instant the body occupies a 
region $\mathcal{B}\subset \R^2$.  The flow is assumed to take place in the connected unbounded
region  $\mathcal{U}:=\R^2\setminus \mathcal{B}$ that is not occupied by the body.  We  assume that  the flow is  potential so the  Eulerian velocity of the fluid $\vecu$ can be written as $\vecu =\nabla \Phi$ for
a  fluid potential $\Phi:\mathcal{U} \to \R$. Incompressibility of the fluid implies that 
$\Phi$ is harmonic,
\begin{equation*}
\label{E:Laplace_Equation}
\nabla^2 \Phi =0 \qquad \mbox{on} \qquad \mathcal{U}.
\end{equation*}

The boundary conditions for $\Phi$ come from the following considerations. On the one hand it is assumed that, up to a purely circulatory flow around the
body,  the motion of the
fluid is solely due to the motion of the body. This assumption   requires the fluid velocity 
$\nabla \Phi $ to vanish at infinity. Secondly, to avoid cavitation or penetration of the fluid into the body,
we require the normal component of the fluid velocity at a material point $p$ on the boundary of $\mathcal{B}$ 
to agree with the
normal component of the velocity of $p$. Suppose that the vector $( X,Y) \in \R^2$ gives   body coordinates for $p$. The 
latter boundary condition is expressed as
\begin{equation*}
\label{E:Boundary_conditions}
\left .  \frac{\partial \Phi}{\partial n} \right |_{p\in \partial \mathcal{B}} =  (v_1-\omega Y)n_1
 +(v_2 +\omega X)n_2,
\end{equation*}
where  ${\bf n}=(n_1,n_2)$ is the outward unit  normal vector to $\mathcal{B}$ at  $p$ written
in body coordinates. These conditions determine the flow of the fluid up to a purely 
circulatory flow around the body  that would persist if the body
is brought to rest. The latter is specified by the value of the circulation $\kappa$ around the body as we
now discuss.

The potential $\Phi$ that satisfies the above boundary value problem  can be
written in terms of the body's  velocities $\omega, v_1, v_2$, in \emph{Kirchhoff form}:
  \begin{equation}
\label{E:Kirch_Potential2D}
\Phi=  \omega \chi + v_1 \phi_1 +  v_2 \phi_2 + \phi_0,
\end{equation}
where $\phi_i$, $i=0,1,2$, and $\chi$ are harmonic functions on $\mathcal{U}$ whose gradients vanish at infinity and  satisfy:
\begin{equation*}
\label{E:Boundary_conditions_Kirch_form}
\left . \frac{\partial \phi_i}{\partial n} \right |_{\partial \mathcal{B}} =n_i, \; i=1,2, \qquad \left . \frac{\partial \chi} {\partial n} \right  |_{\partial \mathcal{B}} = Xn_2-Yn_1, \qquad \left . \frac{\partial \phi_0}{\partial n} \right |_{\partial \mathcal{B}} =0.
\end{equation*}
The potential $\phi_0$ is multi-valued and  defines the 
circulatory flow around the body.
 The circulation $\kappa$ of the fluid around the body satisfies
\begin{equation}
\label{E:defcirc}
\kappa = \oint_{\partial \mathcal{B}} \vecu \cdot d{\mathbf{l}}= \oint_{\partial \mathcal{B}} \nabla \phi_0 \cdot d{\mathbf{l}},
\end{equation}
and remains constant during the motion.

\paragraph{The total kinetic energy of the fluid-body system.}
Disregarding the  circulatory motion, the kinetic energy of the fluid  is given by
\begin{equation*}
T_\mathcal{F}=\frac{\rho}{2} \int_\mathcal{U}
 || \nabla (\Phi-\phi_0) ||^2 \, dA,
\end{equation*}
where $dA$ is the area element in $\R^2$ and  $\rho$ is the (constant) fluid density.  We have subtracted the circulatory part from the velocity potential, as it is known to give rise to an infinite contribution to the fluid kinetic energy and needs to be regularized away, as in \cite{La1945, Mi1968, Sa1992}.

By substituting  \eqref{E:Kirch_Potential2D} into the above, one can express $T_\mathcal{F}$
 as the quadratic form 
\begin{equation}
\label{E:Fluid_Energy_in_terms_of_velocities}
T_\mathcal{F}= \frac{1}{2} \left ( \sum_{i,j=1}^2 \M^{ij}_\mathcal{F} v_{i}v_j + 2 \sum_{i=1}^2\Jn^{i}_\mathcal{F}v_i\omega +\In_\mathcal{F} \omega^2  \right ),
\end{equation}
where  $\mathcal{M}_\mathcal{F}^{ij}, \Jn_\mathcal{F}^{i}$, $i,j=1,2$, and $\In_\mathcal{F}$  are certain constants that only depend on the body shape. 
Explicitly one has (see \cite{La1945} for details),
\begin{equation*}
\begin{split}
& \M^{ij}_\mathcal{F}=-\rho \int_{\partial \mathcal{B}} \phi_i \frac{\partial \phi_j}{\partial n} \, dl=-\rho \int_{\partial \mathcal{B}} \phi_j \frac{\partial \phi_i}{\partial n} \, dl , \; i,j=1,2,  \qquad \In_\mathcal{F}=-\rho \int_{\partial \mathcal{B}} \chi \frac{\partial \chi}{\partial n} \, dl
 \\ & \Jn^i_\mathcal{F}=-\rho \int_{\partial \mathcal{B}} \phi_i \frac{\partial \chi}{\partial n} \, dl = -\rho \int_{\partial \mathcal{B}} \chi \frac{\partial \phi_i}{\partial n} \, dl, \; i=1,2.\end{split}
\end{equation*}
These constants are referred to as \emph{added masses} and are conveniently written
in $3\times 3$ matrix form to define the (symmetric) \emph{added inertia tensor}:
\begin{equation*}
\I_\mathcal{F}:=\left ( \begin{array}{cc} \In_\mathcal{F} & \Jn_\mathcal{F} \\ \Jn_\mathcal{F}^T  & \mathcal{M}_\mathcal{F} \end{array} \right ),
\end{equation*}
that defines $T_\mathcal{F}$ as a quadratic form on $\se(2)$.

On the other hand, the kinetic energy of the body is given by 
\begin{equation*}
\label{E:Body_Energy_in_terms_of_velocities}
T_\mathcal{B}= \frac{1}{2} \left ( (\mathcal{I} + m(a^2 +b^2))\omega^2 + mv_1^2 +mv_2^2-mb\omega v_1 
+ma \omega v_2 \right ),
\end{equation*}
where $m$ is the total mass of the body and $\mathcal{I}$ is its moment of inertia of the body about its
center of mass. We can write $T_\mathcal{B}$ as a quadratic form on $\se(2)$ with matrix
\begin{equation*}
\I_\mathcal{B}:=\left ( \begin{array}{ccc} \In + m(a^2 +b^2) & -mb & ma \\ -mb  & m & 0 \\
ma & 0 & m \end{array} \right ).
\end{equation*}
We define $\I:=\I_\mathcal{F} + \I_\mathcal{B}$ as the {\em total inertia tensor} of the system. We consider
it as an operator $\I : \se(2) \to \se(2)^*$ that satisfies that the left invariant Lagrangian
defined at the identity by
 $$L(\xi) =\frac{1}{2}\langle \I \xi, \xi \rangle,$$ 
 is the
total kinetic energy of the fluid-body system.

\subsection{The Kirchhoff and Chaplygin-Lamb equations}
\label{sec:dynamics}

\paragraph{The Kirchhoff equations.} The (reduced) equations of motion for the motion of a planar body in a potential flow, in the absence
of circulation are the well-known Kirchhoff equations
\begin{equation}
\label{E:Kirchhoff_2D}
\begin{split}
\dot k &= v_2p_1-v_1p_2, \\
\dot p_1 &= \omega p_2, \quad  \quad
\dot p_2 = - \omega p_1.
\end{split}
\end{equation}
Here $k$ and $(p_1,p_2)$ are known as ``impulsive pair" and ``impulsive force" respectively.
They are linearly related to  the body's  velocities $\omega, v_1, v_2$ via the total inertia tensor $\I$.
The above equations are readily shown to be Hamiltonian with respect to the Lie-Poisson bracket 
\eqref{E:Lie-Poisson-se(2)} and the Hamiltonian function $H:\se(2)^*\rightarrow \R$  given by
\begin{equation}
\label{E:Hamiltonian}
H(\mu)=\frac{1}{2}\mu \I^{-1}\mu^T.
\end{equation}

\paragraph{The Chaplygin-Lamb equations.} In the presence of circulation, the Kirchhoff equations on $\se(2)^*$ have to be modified to include the Kutta-Zhukowski force.  This is a gyroscopic force, which is proportional to the circulation $\kappa$.  In this case, the equations of motion are referred to as the \emph{Chaplygin-Lamb equations} and they are given by 
\begin{equation}
\label{E:Kirchhoff_1_2D_circulation}
\begin{split}
\dot k &= v_2p_1-v_1p_2 - \rho(\alpha v_1 + \beta v_2), \\
\dot p_1 &= \omega p_2 - \kappa \rho v_2 + \rho \alpha \omega, \\
\dot p_2 &= - \omega p_1 + \kappa \rho v_1 + \rho \beta \omega,
\end{split}
\end{equation}
where 
the constants $\alpha$ and $\beta$ are  proportional to the circulation $\kappa$ and depend on the position and orientation of the 
body axes. They are explicitly given by:
\begin{equation}
\label{E:circ_constants}
\alpha= \oint_{\partial \B}X\nabla \phi_0 \cdot  \, d\mathbf{l}, \qquad \beta= \oint_{\partial \B}Y\nabla \phi_0 \cdot  \, d\mathbf{l},
\end{equation}
where, as before,  $(X,Y)$ are body coordinates  for material points in $\partial \mathcal{B}$.  The Chaplygin-Lamb equations were first derived in \cite{Ch1933, La1945} and analyzed further in \cite{BoMa2006} (see also the references therein). In  \cite{VaKaMa2010}, the Chaplygin-Lamb equations were derived by considering the interaction between a rigid body and a potential flow with circulation, using techniques from symplectic reduction theory.

\begin{remark} \label{Rmk:Shift} One easily verifies that if the center of the body axes is displaced to
the point with body coordinates $(r,s)$, so that the new body coordinates are 
$\tilde X=X-r, \; \tilde Y=Y-s$, then the circulation  constants relative to the new coordinate axes take
the form $\tilde \alpha=\alpha -r\kappa, \; \tilde \beta =\beta -s\kappa$. Thus, there is a unique choice 
 of the body axes  that makes these constants vanish. On the other hand,
it is also often desirable to choose the body axes so that the total inertia tensor $\I$ is diagonal.
For an asymmetric body, these two choices are in general incompatible,  see e.g. \cite{La1945}.
\end{remark}

 For our purposes, looking ahead to the introduction of the nonholonomic constraint, it is useful to consider  equations \eqref{E:Kirchhoff_1_2D_circulation} in their full
generality where $\alpha, \, \beta\neq 0$, and $\I$ is not  diagonal. This contrasts with 
the treatment in \cite{VaKaMa2010} where it is assumed that $\alpha=\beta=0$ and
with \cite{BoMa2006} where the complementary assumption, namely that $\I$ is diagonal,
is made.
 
 It is shown in \cite{VaKaMa2010} that, if both circulation constants $\alpha$ and $\beta$ vanish,
 the Chaplygin-Lamb equations \eqref{E:Kirchhoff_1_2D_circulation} can be interpreted as
 Lie-Poisson equations on the oscillator group that is a central extension of $\operatorname{SE}(2)$ by
 $\R$ and will be reviewed in Section  \ref{A:SE(2)-Extensions}. We will give a generalization of this
 result to the case in which $\alpha$ and $\beta$ are arbitrary. To do so,  notice, by a direct calculation, that 
 the equations \eqref{E:Kirchhoff_1_2D_circulation} are Hamiltonian with respect to the usual Hamiltonian $H:\se (2)^*\to \R$ given in \eqref{E:Hamiltonian} and with respect to the following 
 non-canonical bracket of functions
on $\se(2)^*$:
\begin{equation}
\label{E:Lie-Poisson-Magnetic}
	\{ F, K \}_{\kappa,\alpha, \beta} (\mu) = ( \nabla_\mu F)^T 
		\begin{pmatrix}
		    0 & -p_2 & p_1 \\
		    p_2 & 0 & 0 \\
		    -p_1 & 0 & 0 
		\end{pmatrix}
		(\nabla_\mu K)- \begin{pmatrix}
		    \rho \kappa\\ -\rho \beta \\
		    \rho \alpha 
		\end{pmatrix} \cdot \left ( ( \nabla_\mu F) \times
		(\nabla_\mu K) \right ).
\end{equation}
Here ``$\times$" denotes the standard vector product in $\R^3$, and, as before, $\nabla_\mu F$ is the gradient of $F$ with respect to the variables $(k, p_1, p_2) = \mu$. In fact we have:
 \begin{theorem}
\label{T:Lie-Poisson_structure_of_Chap_Lamb}
The Chaplygin--Lamb equations \eqref{E:Kirchhoff_1_2D_circulation} are of Lie-Poisson
type on the  dual Lie algebra $\widehat \g^*=\mathfrak{se}(2)^*\times \R^3$ of a central extension  
of $\operatorname{\operatorname{SE}}(2)$ by $\R^3$. 

\end{theorem}

The proof of this theorem is postponed to Section  \ref{A:SE(2)-Extensions} where we introduce the appropriate
central extension $\widehat G$ of $\operatorname{SE}(2)$. The explicit
formula for the Lie-Poisson bracket on $\widehat \g^*$ is given in Proposition \ref{P:Formula-Lie-Poisson-Ext}.
 The non-canonical bracket  \eqref{E:Lie-Poisson-Magnetic}
on $\se(2)^*$ arises from the Lie-Poisson bracket on $\widehat \g^*$ when one fixes
the value $\boldsymbol{\sigma}=(\rho \kappa, -\rho \beta , \rho \alpha )\in \R^3$ as explained 
in Section \ref{Sec:Central-Extension}, equations \eqref{E:Lie_Poisson_C_Ext}, \eqref{noncanpoisson}.

\subsection{The hydrodynamic Chaplygin sleigh with circulation}
\label{SS:hydro-Chap-sleigh}

We now introduce the nonholonomic constraint.  Recall that the classical Chaplygin sleigh problem (going back to 1911, \cite{Ch2008}) describes the motion of a planar rigid body with a knife edge (a blade) sliding on a horizontal plane. The nonholonomic constraint forbids the motion in the direction perpendicular to the blade. In its hydrodynamic version, the body is surrounded by a potential fluid and the nonholonomic constraint models the effect of a very effective fin or keel, see  \cite{FeGa2010}. 

\begin{figure}[h]
\begin{center}
	\includegraphics[scale=.2]{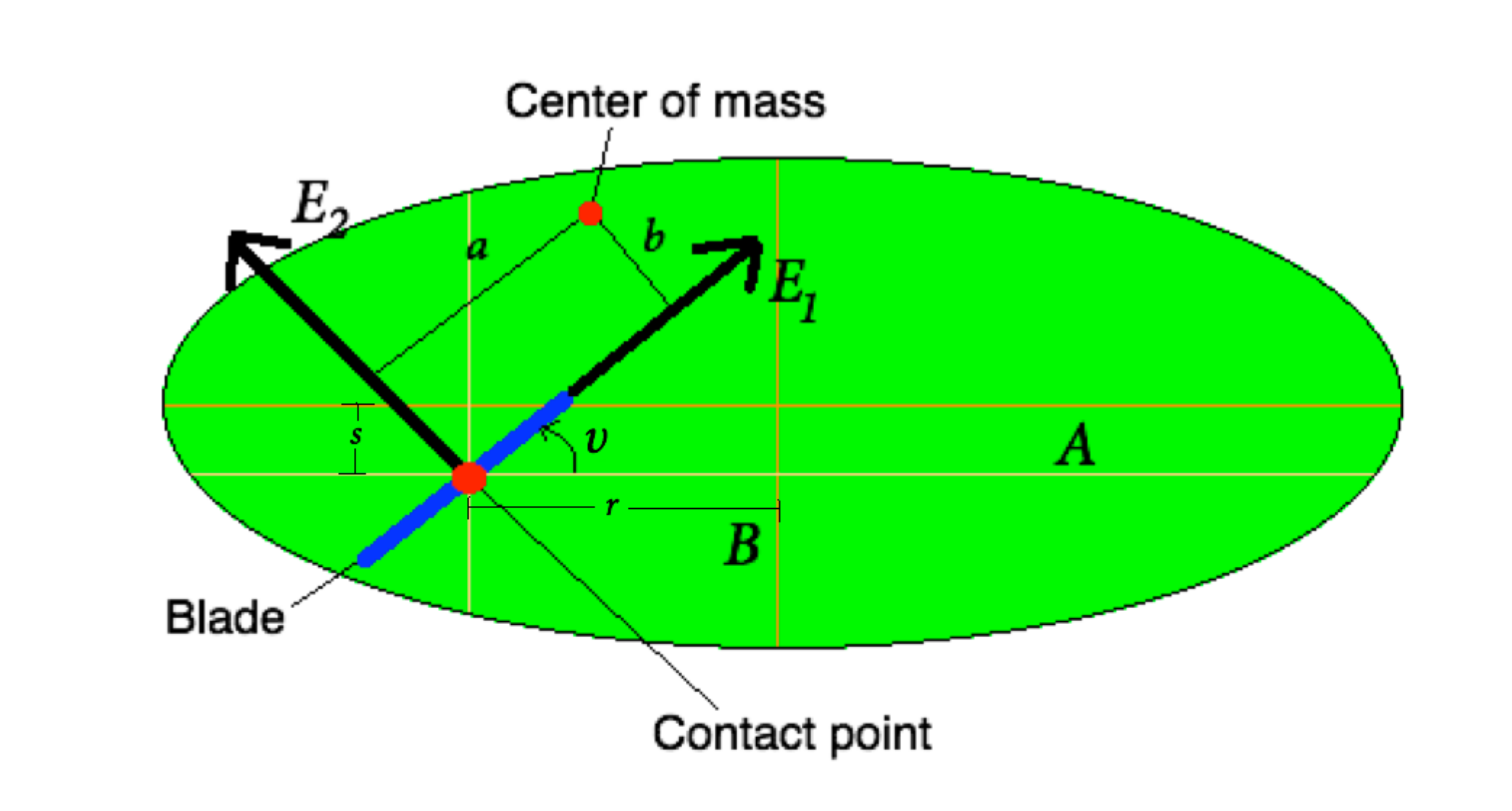}
	\caption{The hydrodynamic Chaplygin sleigh consists of a rigid body moving in a potential fluid equipped with a very effective fin or blade, which prevents motion in the direction lateral to the fin.  The body frame $\{ \mathbf{E}_1, \mathbf{E}_2 \}$ is taken so that $\mathbf{E}_1$ is aligned with the fin. 
 The fin does not have to be aligned with any symmetry axes of the body, and we make no assumptions about the relative location of the center of mass, the geometric center of the body, and the contact point of the fin. 
	We have depicted an elliptical body shape, but our results are valid for arbitrary convex shapes. \label{fig:sleighdiagram} }
\end{center}
\end{figure}

With the notation of section~\ref{sec:kinematics}, we let $\{{\bf E}_1, {\bf E}_2 \}$ be a body frame located at the contact point of the sleigh and the
plane, and so that the ${\bf E}_1$-axis is aligned with the blade (see Figure \ref{fig:sleighdiagram}).  

The resulting nonholonomic constraint is given by $v_2=0$, and is clearly left invariant under the action of  $\operatorname{\operatorname{SE}}(2)$, as it is solely written in terms of the velocity of the body as seen in the body frame.  

In the absence of constraints, the motion of the body  is described by the Chaplygin--Lamb equations \eqref{E:Kirchhoff_1_2D_circulation}
that, as we have shown (Theorem \ref{T:Lie-Poisson_structure_of_Chap_Lamb}), are Lie-Poisson
equations on the dual Lie algebra of the central extension  of $\operatorname{\operatorname{SE}}(2)$ 
by $\R^3$ and with respect to
the  kinetic energy Hamiltonian \eqref{E:Hamiltonian}.

 The reduced nonholonomic
equations are thus of EPS type on  $\widehat \g^*=\se (2)^*\times \R^3$.
Note that the co-vector $(0,0,1)\in \g^*$ annihilates all  elements $\xi=(\omega,v_1,v_2)\in \se(2)$
that satisfy the constraints. Thus, in agreement with the results of Theorem    \ref{T:Lie-Poisson_structure_of_Chap_Lamb},  by  putting $\mu =(k,p_1,p_2)\in \se(2)^*$ 
and $\sigma =(\rho \kappa, -\rho \beta, \rho \alpha)\in \R^3$,   we get the following explicit expression
for the reduced nonholonomic equations \eqref{E:Euler-Poincare-Suslov-Cent-Ext}, 
\begin{equation}
\begin{split}
\label{E:Constrained_2D_circulation}
\dot k &= v_2p_1-v_1p_2 - \rho(\alpha v_1 + \beta v_2), \\
\dot p_1 &= \omega p_2 - \kappa \rho v_2 + \rho \alpha \omega, \\
\dot p_2 &= - \omega p_1 + \kappa \rho v_1 + \rho \beta \omega + \lambda,
\end{split}
\end{equation}
where the multiplier $\lambda$ is determined from the condition $v_2=0$.

\paragraph{Detailed equations of motion.}

In the sequel we assume that the shape of the sleigh is arbitrary convex and that its center of mass  does
not necessarily coincide with the origin, which leads to the general total inertia tensor
\begin{equation*}
\label{E:Total_Inertia_Ellipse_Rot_Axes}
\I= \left ( \begin{array}{ccc} J & -L_2 & L_1 \\ -L_2 & M &Z  \\ L_1& Z & N \end{array} \right ) ,
\end{equation*}
and with arbitrary circulation constants $\alpha, \, \beta$. The particular expressions for $\alpha, \beta$ and
$\I$ in the case that the body is of elliptic shape are given in \cite{FeGaVa2012}.

In accordance with  \eqref{E:Lagrange_multiplier}, the multiplier   $\lambda$ is given by:
\begin{equation*}
\label{E:Lagrange_Multiplier_part}
\lambda=-\frac{1}{( { \I^{-1}} )_{33}} \left( \I^{-1} \left ( \begin{array}{c} v_2 p_1 - v_1 p_2 -\rho(\alpha v_1 + \beta v_2) \\ \omega p_2 -\kappa \rho v_2 + \rho \alpha \omega \\ -\omega p_1 + \kappa \rho v_1 + \rho \beta \omega) \end{array} \right ) \right )_3,
\end{equation*}
where
\begin{equation*}
\I^{-1}= \frac{1}{\mbox{det}(\I)} \left ( \begin{array}{ccc} MN-Z^2 & ZL_1+NL_2 & -ZL_2-ML_1 \\ ZL_1+NL_2 & JN-L_1^2 &-L_1L_2-JZ  \\ -ZL_2-ML_1 & -L_1L_2-JZ  & JM-L_2^2 \end{array} \right ).
\end{equation*}

A long but straightforward calculation shows that,
by expressing $\omega, v_1$ and $v_2$ in terms of  $k, p_1, p_2$, substituting into \eqref{E:Constrained_2D_circulation},
and  enforcing the constraint $v_2=0$, one obtains:
\begin{equation}
\label{E:Working_Hydro_Sleigh_Equations}
\begin{split}
\dot \omega &=\frac{1}{D}\left (L_1 \omega + Z v_1 + \rho \alpha \right  ) \left ( L_2 \omega - Mv_1\right ),  \\
\dot v_1 &=\frac{1}{D} \left (L_1 \omega + Z v_1  + \rho \alpha \right ) \left (J\omega -L_2 v_1 \right ),
\end{split}
\end{equation}
where we set $D= \operatorname{det} (\I)( {\I^{-1}} )_{33}= MJ-L_2^2$. Note that $D>0$ since $\I$ and $\I^{-1}$ are positive definite. 
Note as well that if $\alpha=0$ we recover
 the system with zero circulation treated in \cite{FeGa2010} so from now on we assume $\alpha \neq 0$.

The full motion of the sleigh on the plane is determined by the reconstruction equations 
which,
in our case with $v_2=0$, reduce to
\begin{equation*} \label{recon}
\dot \theta = \omega, \qquad \dot x=v_1\cos \theta, \qquad \dot y = v_1\sin \theta.
\end{equation*}

The reduced energy integral has
\begin{equation*}
H=\frac{1}{2}\left ( J\omega^2 +Mv_1^2 -2L_2\omega v_1 \right ),
\end{equation*}
and its level sets are ellipses on the $(\omega \, v_1)$-plane.

As seen from the equations,
the straight line $\ell=\{L_1 \omega + Z v_1 + \rho  \alpha=0\}$ consists of equilibrium points for the system. Each of these equilibria corresponds to a uniform circular motion on the plane along a 
circumference of radius $\left | \frac{v_1}{\omega} \right |$.

 Notice that if $Z=L_1=0$  the line $\ell$ of equilibra disappears.
In fact, it is shown in \cite{FeGa2010} that in the absence of circulation the system possesses an
invariant measure only for this specific value of the parameters. In view of Theorem \ref{T:inv-meas}
we conclude
\begin{proposition} \label{prop:chaplygin_measure}
The equations of motion \eqref{E:Working_Hydro_Sleigh_Equations} possess an invariant
measure if and only if $Z=L_1=0$.
\end{proposition}

\begin{figure}[h]
\begin{center}
	\includegraphics[scale=.4]{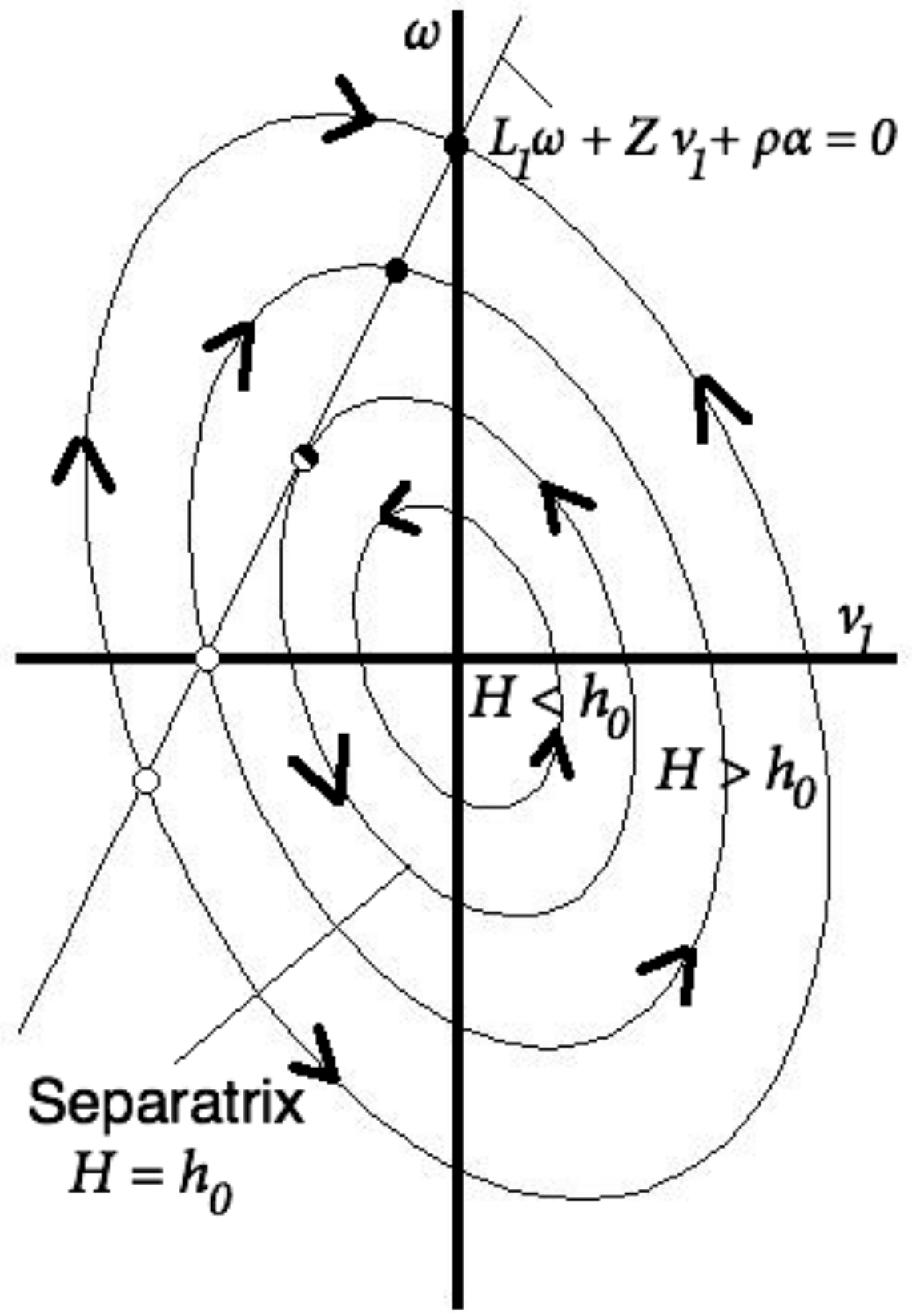} 
	\caption{Reduced phase portrait. The line $\ell$ consists of equilibria that are either stable (filled dots)
	or unstable (empty dots). The trajectories are contained in the level surfaces of the Hamiltonian $H$
	that are ellipses on the $v_1\;\omega$ plane. \label{F:phase_space}}
\end{center}
\end{figure}

In this  case we obtain simple harmonic motion on the reduced plane $\omega, v_1$.
The reduced phase space in the case where $Z\ne 0$ and $L_1 \ne 0$ is illustrated in Figure~\ref{F:phase_space}. As shown in \cite{FeGaVa2012}, in this case there exists a positive value of the energy $h_0$ that divides periodic from heteroclinic orbits. The separatrix corresponding to $H=h_0$ is a homoclinic 
orbit. 

For any value of the parameters the reduced system (\ref{E:Working_Hydro_Sleigh_Equations}) can be checked to be Hamiltonian
with respect to the following Poisson bracket of functions of $\omega, v_1$ that can be obtained
using the results of \cite{Ga2007}
\begin{equation*}
\{F_1,F_2 \}:= -\frac{1}{D}\left (L_1 \omega + Z v_1 + \rho \alpha \right ) \left ( \frac{\partial F_1}{\partial \omega} \frac{\partial F_2}{\partial v_1} -
\frac{\partial F_1}{\partial v_1} \frac{\partial F_2}{\partial \omega} \right ).
\end{equation*}
The invariant symplectic leaves  consist of the semi-planes separated by the equilibria line $\ell$
and the zero-dimensional leaves formed by the points on this line.

The integration of the reduced equations as well as a detailed study of the motion of the sleigh on the 
plane is given in \cite{FeGaVa2012}.

\section{Central extensions of the euclidean group $\operatorname{SE}(2)$}
\label{A:SE(2)-Extensions}

The purpose of this section is to explain in detail the central extension of $\operatorname{SE}(2)$
that encodes the effects of circulation in the Chaplygin-Lamb equations and appears in 
the statement of Theorem \ref{T:Lie-Poisson_structure_of_Chap_Lamb} and allows for the
 construction given in section \ref{SS:hydro-Chap-sleigh}.

\subsection{The oscillator group}

We start by defining the real valued $\operatorname{\operatorname{SE}}(2)$-two-cocycle $B_1:\operatorname{\operatorname{SE}}(2)\times \operatorname{\operatorname{SE}}(2) \to \R$
by
\begin{equation} \label{cocycle1}
B_1((R_\theta, {\bf x} ),(R_{\theta'} , {\bf x}'))=\frac{1}{2}{\bf x}\cdot \mathbb{J} R_\theta  {\bf x}',
\end{equation}
where $\mathbb{J}$ is the $2\times 2$ symplectic matrix
\begin{equation*}
\mathbb{J}=\left ( \begin{array}{cc} 0 & 1 \\ -1 & 0 \end{array} \right ).
\end{equation*}
This cocycle differs from the one used in \cite{VaKaMa2010} by a multiplicative factor of $\frac{1}{2}$.  On the Lie algebra level, we have by \eqref{E:Lie_algebra_cocycle} that the infinitesimal cocycle $C_1:\mathfrak{se}(2)\times \mathfrak{se}(2) \to \R$ is given by 
\[
C_1\left ( (\omega, v_1,v_2), (\omega', v_1',v_2') \right ) = v_1v_2'-v_2v_1'.
\]

The central extension of $\operatorname{SE}(2)$ by $\mathbb{R}$ using the cocycle $B_1$ is referred to as the \emph{oscillator group} \cite{St1967}, and will be denoted as $\mathrm{Osc}$.  The Lie algebra $\mathfrak{osc}$ of the oscillator group is isomorphic to $\se(2) \times \mathbb{R}$, with Lie bracket
\begin{align*}
\left	[ \, (\omega, v_1,v_2; z ) \, , \, (\omega', v_1',v_2' ; z') \, \right  ]_{\mathfrak{osc}} & = 
\left (	  \left [ \, (\omega, v_1,v_2) \, , \,  (\omega', v_1',v_2')\, \right ]_{\mathfrak{se}(2)}  \, ; \, C_1\left ( (\omega, v_1,v_2), (\omega', v_1',v_2') \right )  \right ) \\
	 & = ( \, 0 \, , \,  v_2\omega'-\omega v_2' \, , \,  \omega v_1' -v_1 \omega ' \, ;  \, v_1v_2'-v_2v_1' \, ).
\end{align*}
On the dual $\mathfrak{osc}^\ast \cong \se(2)^\ast \times \mathbb{R}^\ast$, the (minus)
 Lie-Poisson bracket is given by \eqref{E:Lie_Poisson_C_Ext}, or explicitly by 
\begin{equation} \label{oscpoisson}
	\{ F, K \}^-_{\mathfrak{osc}^\ast}(\mu, \sigma_0) = \{ F, K \}^-_{\se(2)^\ast}(\mu) - \sigma_0 
	C_1 \left( \frac{\delta F}{\delta \mu}, \frac{\delta K}{\delta \mu} \right),
\end{equation}
for $(\mu, \sigma_0) \in \se(2)^\ast \times \mathbb{R}^\ast$.  It is easy to see that this bracket
coincides with the bracket \eqref{E:Lie-Poisson-Magnetic} in the case where the circulation constants $\alpha$ and $\beta$
vanish, and where $\sigma_0$ plays the role of the circulation $\rho \kappa$.  We conclude that,
modulo the presence of non-zero circulation constants $\alpha$ and $\beta$, the effect of non-zero circulation can be described in terms of the geometry of the oscillator group, and more precisely the cocycle \eqref{cocycle1}. This had already been established in \cite{VaKaMa2010}.

In terms of Lie algebra cohomology, it is easy to show that $C_1$ is a closed two-cocycle which is not exact.  Furthermore, as $H^2( \se(2), \mathbb{R})$ is isomorphic to $\mathbb{R}$, we have that $C_1$ determines a generator of the second cohomology.   Since isomorphy classes of central extensions are classified by the second Lie algebra cohomology, we may refer to the oscillator group as \emph{the} central extension of $\operatorname{SE}(2)$ by $\mathbb{R}$.

\subsection{A central extension of $\operatorname{\operatorname{SE}}(2)$ by $\R^3$.}

We now describe the effect of nonzero $\alpha$ and $\beta$ in the Poisson bracket \eqref{E:Lie-Poisson-Magnetic}.  To this end, we observe that the equations of motion \eqref{E:Kirchhoff_1_2D_circulation} for nonzero $\alpha$ and $\beta$ can be obtained from the equations where $\alpha = \beta = 0$ by making the substitution, or \emph{momentum shift},
\begin{equation} \label{momentumshift}
	p_1 \leadsto p_1 + \sigma_1, \quad p_2 \leadsto p_2 + \sigma_2,
\end{equation}
where $\sigma_1 = - \rho \beta$ and $\sigma_2 = \rho \alpha$, 
while the angular momentum $k$ remains invariant.  This observation seems natural in view of 
Remark \ref{Rmk:Shift}.

 The momentum shift \eqref{momentumshift} can be described in geometric terms as follows.  Let $\mathcal{A}: \mathfrak{se}(2) \to \mathbb{R}^3$ be the linear map $\mathcal{A}(\omega, v_1, v_2) = (0, v_1, v_2)$. This map is a one-cocycle on $\se(2)$ with values in $\mathbb{R}^3$ and its derivative is given by 
\begin{align*}
	\delta \mathcal{A}( (\omega, v_1, v_2), (\omega', v_1', v_2') )  =
	- \mathcal{A}([ (\omega, v_1, v_2), (\omega', v_1', v_2')]_{\mathfrak{se}(2)}) 
	= ( \, 0 \, , \, \omega v_2'- v_2\omega' \, , \, v_1 \omega '- \omega v_1'  \,). 
\end{align*}

Now, consider fixed values $\sigma_1, \sigma_2 \in \mathbb{R}$ and define the linear map $\mathcal{A}_{(\sigma_1, \sigma_2)} : \se(2) \to \mathbb{R}$ given by 
\[
	\mathcal{A}_{(\sigma_1, \sigma_2)} (\omega, v_1, v_2) := 
		\left<(0, \sigma_1, \sigma_2), \mathcal{A}(\omega, v_1, v_2) \right> 
		= \sigma_1 v_1 + \sigma_2 v_2.
\]
As $\mathcal{A}_{(\sigma_1, \sigma_2)}$ is an element of $\se(2)^\ast$, we may write the momentum shift \eqref{momentumshift} more formally as the map $\Phi_{\mathcal{A}}: \se(2)^\ast \to \se(2)^\ast$ given by
\begin{equation} \label{geommomentumshift}
	\Phi_{\mathcal{A}}( \mu ) = \mu - \mathcal{A}_{(\sigma_1, \sigma_2)},
\end{equation}
for all $\mu \in \se(2)^\ast$.  The minus sign is due to the fact that this is the \emph{active} version of the transformation \eqref{momentumshift}: if we denote the new momenta of the system by $\bar{p}_1, \bar{p}_2$, then the effect of performing the substitution \eqref{momentumshift} is that the old and the new momenta are related by $p_1 = \bar{p}_1 + \sigma_1, p_2 = \bar{p}_2 + \sigma_2$, which is just \eqref{geommomentumshift}.

As the dual of the Lie algebra of the oscillator group is just the Cartesian product $\se(2)^\ast \times \mathbb{R}^\ast$, the map $\Phi_{\mathcal{A}}$ gives rise to a map $(\mu, \sigma_0) \mapsto (\Phi_{\mathcal{A}}(\mu), \sigma_0)$ on $\mathfrak{osc}^\ast$ which we denote by $\Phi_{\mathcal{A}}$ as well. We now investigate the behavior of the Poisson structure \eqref{oscpoisson} under the map $\Phi_{\mathcal{A}}$.  Since $\Phi_{\mathcal{A}}$ is a constant shift map, we have for arbitrary functions $F, K$ on $\mathfrak{osc}^\ast$ that 
\[
 \frac{\delta (F \circ \Phi_{\mathcal{A}}^{-1})}{\delta \mu} = 
 	\frac{\delta F}{\delta \mu},
\]
and similarly for $K$, so that $\{ F \circ \Phi_{\mathcal{A}}^{-1}, K \circ \Phi_{\mathcal{A}}^{-1} \}^-_{\mathfrak{osc}^\ast}(\mu, \sigma) = \{ F, K \}^-_{\mathfrak{osc}^\ast}(\mu, \sigma)$.  On the other hand,
\begin{align}
	\{ F, K \}^-_{\mathfrak{osc}^\ast}(\Phi_{\mathcal{A}}^{-1}(\mu), \sigma) & = \{ F, K \}^-_{\se(2)^\ast}(\Phi_{\mathcal{A}}^{-1}(\mu)) - \sigma_0 
	C_1 \left( \frac{\delta F}{\delta \mu}, \frac{\delta K}{\delta \mu} \right)\nonumber \\
	& = \{ F, K \}^-_{\se(2)^\ast}(\mu) -
		\left< \mathcal{A}_{(\sigma_1, \sigma_2)}, 
			\left[ \frac{\delta F}{\delta \mu}, \frac{\delta K}{\delta \mu} 
				\right] \right>
	-
	 \sigma_0 
	C_1 \left( \frac{\delta F}{\delta \mu}, \frac{\delta K}{\delta \mu} \right) \nonumber \\
	& = \{ F, K \}^-_{\se(2)^\ast}(\mu)
	- \left( \sigma_0 C_1 - \delta \mathcal{A}_{(\sigma_1, \sigma_2)}\right)
	\left( \frac{\delta F}{\delta \mu}, \frac{\delta K}{\delta \mu} \right), \label{lhspoisson}
\end{align} 
so that $\Phi_{\mathcal{A}}$ takes the Lie-Poisson bracket on $\mathfrak{osc}^\ast$ into the bracket \eqref{lhspoisson} modified by a cocycle.  The cocycle $\sigma_0 C_1 - \delta \mathcal{A}_{(\sigma_1, \sigma_2)}$ can be made more explicit by noting that 
\begin{align}
	(\sigma_0 C_1 - \delta \mathcal{A}_{(\sigma_1, \sigma_2)})((\omega, v_1, v_2), (\omega', v'_1, v'_2)) 
		 =  \begin{bmatrix} \sigma_0 \\  \sigma_1 \\  \sigma_2 \end{bmatrix} \cdot  \begin{bmatrix}
				v_1 v_2' - v_2 v_1' \\
				v_2 \omega '  - \omega v_2' \\
			 \omega v_1'  - v_1 \omega '
			\end{bmatrix}  
		 = \boldsymbol{\sigma} \cdot \left(  \begin{bmatrix} 
				\omega \\ v_1 \\ v_2
			\end{bmatrix}
				\times
			\begin{bmatrix} 
				\omega ' \\ v_1' \\ v_2'
			\end{bmatrix}	
		 \right), \label{bigcocycle} 
\end{align}
where $\boldsymbol{\sigma}:=(\sigma_0,\sigma_1, \sigma_2)$.
Note that this is precisely the cocycle in the Poisson bracket \eqref{E:Lie-Poisson-Magnetic}
for the specific value of $\boldsymbol{\sigma}:=(\rho \kappa, -\rho \beta, \rho \alpha)$.  We conclude that the effect of introducing nonzero $\sigma_1, \sigma_2$ is to modify the Lie-Poisson bracket on $\mathfrak{osc}^\ast$ by a trivial $\mathbb{R}^2$-valued cocycle $- \delta \mathcal{A}_{(\sigma_1, \sigma_2)}$.   We now let $C_2$ be equal to $- \delta \mathcal{A}$: this is an $\mathbb{R}^3$-valued cocycle on $\se(2)$, which can be integrated to a group two-cocycle $B_2$ on $\operatorname{SE}(2)$, given by
\begin{equation} \label{trivgroupcocycle}
B_2((R_\theta, {\bf x} ),(R_{\theta '} , {\bf x}'))= (0, R_\theta  {\bf x}' -{\bf x}').
\end{equation}

As we showed previously, we can describe the cocycle bracket \eqref{lhspoisson} by considering the central extension of the oscillator group by the cocycle $B_2$ given in \eqref{trivgroupcocycle}.  This is equivalent to extending $\operatorname{SE}(2)$ by $\mathbb{R}^3$ using the combined cocycle $B := (B_1, B_2) : \operatorname{SE}(2) \times \operatorname{SE}(2) \to \mathbb{R}^3$; the result is an extension $\widehat{G}$ which is isomorphic to $\operatorname{SE}(2) \times \mathbb{R}^3$ with multiplication 
\[
	(R_\theta, \mathbf{x}; \mathbf{w}) \cdot (R_{\theta '}, \mathbf{x}'; \mathbf{w}') = 
		(R_{\theta + \theta '}, \mathbf{x} + R_\theta \mathbf{x} ' ; \mathbf{w} + \mathbf{w}' + B( (R_\theta, \mathbf{x}), (R_{\theta '}, \mathbf{x}'))).
\]
The infinitesimal cocycle $C: \se(2) \times \se(2) \to \mathbb{R}^3$ associated to $B$ is given by $C = (C_1, C_2)$, or 
\begin{equation}
\label{E:Circulation_algebra_cocycle}
\begin{split}
C((\omega, v_1, v_2), (\omega', v_1', v_2'))&=(v_1v_2'-v_2v_1',  v_2 \omega' - \omega v_2',
 \omega v_1' -v_1 \omega') \\
&=(\omega, v_1, v_2)\times (\omega', v_1', v_2'),
\end{split}
\end{equation}
which is precisely the cocycle appearing on the right-hand side of \eqref{bigcocycle}.  The bracket on the algebra $\widehat{\mathfrak{g}}:= \mathrm{Lie} ( \widehat G)$ is then given by
\begin{equation*}
\begin{split}
[(\omega, v_1, v_2 ; {\bf z}) , (\omega', v_1', v'_2 ; {\bf z}')]_{\widehat{\mathfrak{g}}}&
=
\left (	  \left [ \, (\omega, v_1,v_2) \, , \,  (\omega', v_1',v_2')\, \right ]_{\mathfrak{se}(2)}  \, ; \, C \left ( (\omega, v_1,v_2), (\omega', v_1',v_2') \right )  \right ) \\
	 & = \left ( \, 0 \, , \,  v_2\omega'-\omega v_2' \, , \,  \omega v_1' -v_1 \omega ' \, ;  \, v_1v_2'-v_2v_1', 
	  v_2 \omega' - \omega v_2',  \omega v_1' -v_1 \omega' \, \right ).
\end{split}
\end{equation*}

\paragraph{The Lie-Poisson bracket on $\widehat{\mathfrak{g}}^*$.} We write explicitly the (minus) Lie-Poisson bracket on 
the dual Lie algebra $\widehat{\mathfrak{g}}^*$. Note first that as a  vector space $\widehat{\mathfrak{g}}^*$ is just
$\mathfrak{se}(2)^*\times \R^3$, so that an element $\nu$ of $\widehat{\mathfrak{g}}^*$ can be written as
$\nu:=(\mu, \boldsymbol{\sigma})$ where $\mu\in \mathfrak{se}(2)^*$ and $\boldsymbol{\sigma} \in \R^3$.
In view of \eqref{E:Lie_Poisson_C_Ext} and \eqref{E:Circulation_algebra_cocycle} we obtain:
\begin{proposition}
\label{P:Formula-Lie-Poisson-Ext}
The (minus) Lie-Poisson bracket of functions $F,K\in C^\infty(\widehat{\mathfrak{g}}^*)$ is given by
\begin{equation*}
\{ F, K\}^-_{\widehat{\mathfrak{g}}^*}(\mu, \boldsymbol{\sigma})=\{ F, K\}^-_{{\mathfrak{se}(2)}^*}(\mu)-\boldsymbol{\sigma}\cdot 
\left( \frac{\delta F}{\delta \mu} \times \frac{\delta K}{\delta \mu} \right),
\end{equation*}
where we identify $\mathfrak{se}(2)$ with $\R^3$ in the usual way to make sense of the vector product
of the functional derivatives.
\end{proposition}

In coordinates, the Poisson structure of the above proposition is
\begin{equation*}
\{ F, K\}^-_{\widehat{\mathfrak{g}}^*}(\mu, \boldsymbol{\sigma})= (\nabla_{(\mu, \boldsymbol{\sigma})}F )^T \, \left ( \begin{array}{cccccc}
0 & -p_2 -\sigma_2 & p_1 +\sigma_1 & 0 &0 & 0 \\
p_2+\sigma_2& 0  & -\sigma_0 & 0 & 0 & 0 \\
-p_1 -\sigma_1& \sigma_0 & 0 & 0 & 0 & 0 \\
0 & 0 & 0 & 0 & 0 &0 \\
0 & 0 & 0 & 0 & 0 &0 \\
0 & 0 & 0 & 0 & 0 &0  \end{array} \right )\, (\nabla_{(\mu, \boldsymbol{\sigma})}K),
\end{equation*}
where as usual $\mu=(k,p_1,p_2)\in \se(2)^*$ and $\boldsymbol{\sigma}=(\sigma_0,\sigma_1,\sigma_2)\in \R^3$.

\paragraph{Proof of Theorem \ref{T:Lie-Poisson_structure_of_Chap_Lamb}.} \begin{proof} The reduced kinetic energy Hamiltonian $H(\mu)=\frac{1}{2}\mu \cdot \I^{-1} \mu$ on $\se(2)^*$
defined  in \eqref{E:Hamiltonian} is naturally interpreted as a Hamiltonian on $\widehat{\mathfrak{g}}^*$.
The equations of motion for this Hamiltonian
 and the bracket $\{ \cdot, \cdot \}^-_{\widehat{\mathfrak{g}}^*}$ are then given in coordinates by
\begin{equation*}
\begin{split}
\dot k&= -p_2v_1+p_1v_2 -\sigma_2v_1+\sigma_1v_2, \\
\dot p_1&=p_2\omega + \sigma_2\omega- \sigma_0v_2, \\
\dot p_2&=-p_1\omega-\sigma_1\omega+\sigma_0v_1, \\
\dot {\boldsymbol{\sigma}}&=0,
\end{split}
\end{equation*}
that coincide with  \eqref{E:Kirchhoff_1_2D_circulation} if we put $\sigma_0=\rho \kappa, \; \sigma_1=-\rho \beta, \;
\sigma_2= \rho \alpha$.  \end{proof}

Finally, we mention that  the above equations possess the conserved quantity
\begin{equation*}
\bar F(\mu,\boldsymbol{\sigma})= p_1^2+p_2^2+2\sigma_0k+2\sigma_1p_1+2\sigma_2p_2,
\end{equation*}
which is a Casimir function of the bracket $\{ \cdot, \cdot \}^-_{\widehat{\mathfrak{g}}^*}$. For a  fixed value of $\boldsymbol{\sigma}$, the regular level sets of $\bar F$ define a  symplectic
foliation of $\se(2)^*$. It is easily seen that the leaves of such foliation  are 
paraboloids if $\sigma_0\neq 0$, and
cylinders otherwise. The trajectories of the system are contained in the intersection of the level sets 
of $\bar F$ and $H$.

\section*{Future work}

For the future, we intend to study the motion of the hydrodynamic Chaplygin sleigh coupled to {\em point vortices} in the fluid \cite{Ne2001}.   The equations of motion for interacting point vortices and rigid bodies (without nonholonomic constraints) were recently derived in \cite{ShMaBuKe2002, BoMaRa2003} and since then there have been significant efforts towards discerning integrability and chaoticity \cite{Ra2001, RoAr2010} and towards uncovering the underlying geometry of these models \cite{VaKaMa2009}.  We plan on coupling the nonholonomic Chaplygin sleigh with one or several point vortices in the flow, taking these models as our starting point. We hope to report with progress on these problems  in the near future.

{\small
\paragraph{Acknowledgments.} 
We thank the GMC (Geometry, Mechanics and Control Network, project
MTM2009-08166-E, Spain) for facilitating our collaboration during the events that it organizes. 
We thank Yuri Fedorov for his valuable suggestions and remarks. We are also
thankful to Hassan Aref, Larry Bates, Paul Newton and J{\polhk{e}}drzej {\'S}niatycki
 for useful and interesting discussions.
JV is partially supported by the {\sc irses} project {\sc
geomech} (nr.\ 246981) within the 7th European Community Framework Programme, and is on leave from a Postdoctoral Fellowship of the Research Foundation--Flanders (FWO-Vlaanderen).
LGN thanks the hospitality of the Dipartimento de Matematica Pura e Applicata of University of Padova where part of  this
work was done.
}

\end{document}